\documentclass[reprint,superscriptaddress,amsmath,amssymb,aps,prl,floatfix]{revtex4-2}
\usepackage{graphicx} 
\usepackage[caption=false]{subfig}
\usepackage{epsfig} 
\usepackage{bm}
\usepackage{amssymb}
\usepackage{amsmath}
\usepackage{xcolor}

\begin{document}


\title{Wetting controlled boiling at the nanoscale} 
\author{Oscar Guti\'errez-Varela}
\affiliation{Instituto de F\'isica, Universidad Nacional Aut\'onoma de M\'exico, Ciudad de M\'exico, M\'exico}
\affiliation{Univ. Lyon, Universit\'e Claude Bernard Lyon 1, CNRS, Institut Lumière Matière, F-69622, Villeurbanne, France}

\author{Julien Lombard}
\affiliation{Departamento de F\'isica y Qu\'imica Te\'orica and Departamento de Matem\'aticas, Facultad de Qu\'imica, Universidad Nacional Aut\'onoma de M\'exico, Ciudad de M\'exico, M\'exico}

\author{Thierry Biben}
\affiliation{Univ. Lyon, Universit\'e Claude Bernard Lyon 1, CNRS, Institut Lumière Matière, F-69622, Villeurbanne, France}

\author{Ruben Santamaria}
\affiliation{Instituto de F\'isica, Universidad Nacional Aut\'onoma de M\'exico, Ciudad de M\'exico, M\'exico}

\author{Samy Merabia}
\affiliation{Univ. Lyon, Universit\'e Claude Bernard Lyon 1, CNRS, Institut Lumière Matière, F-69622, Villeurbanne, France}

\date{\today}
             
\begin{abstract}
  Boiling is the out-of-equilibrium transition which occurs when a liquid is heated above its vaporization temperature. At the nanoscale, boiling may be triggered by irradiated nanoparticles immersed in water or nanocomposite surfaces and often results in micro-explosions. It is generally believed that nanoscale boiling occurs homogeneously when the local fluid temperature exceeds its spinodal temperature, around $573$ K for water.
Here, we employ molecular dynamics simulations to show that nanoscale boiling is an heterogenous phenomenon occuring when water temperature exceeds a wetting dependent onset temperature. This temperature can be $100$ K below spinodal temperature if the solid surface is weakly wetting water.  In addition, we show that boiling is a slow process controlled by the solid-liquid interfacial thermal conductance, which turns out to decrease significantly prior to phase change yielding long nucleation times. We illustrate the generality of this conclusion by considering both a heated metallic nanoparticle immersed in water and a solid surface displaying nanoscale wetting heterogeneities. These results pave the way to control boiling using nanoscale patterned surfaces. 
\end{abstract}


\maketitle

Boiling, the transition from a liquid state to a vapor state driven by a heated solid, may proceed along two different scenarios: normal boiling occurs when a vapor nucleus is formed and detaches from the solid. For water, this typically corresponds to a local temperature $\sim 373$ K. Explosive boiling occurs when a fluid is heated up so fast that it may remain trapped in its metastable state until it crosses the liquid-vapor spinodal line, around $573$ K for water. While normal boiling is common to gentle heating, explosive boiling is associated with abrupt heating, which is relevant in the volcanoe explosions and nuclear power plants accidents~\cite{sheperd1982}.

Boiling may be also investigated at the nanoscale. Plasmonic nanoparticles dispersed in water, susceptible to be heated up on a picosecond time scale by a laser, act as local hot spots in a liquid environment
and constitute a platform to probe the physics of phase transitions at the nanoscale~\cite{kotaidis2006,hou2015,wang2018,sasikumar2014}. 
From a practical perspective, the possibility to convert light into vapor has been exploited in several scientific areas, including nanoparticle assisted cancer therapy \cite{shao2015,pitsillides2003,kitz2011}, micro and nano manipulation \cite{zhao2014}, and photoacoustic imaging~\cite{lapotko2009-1,zharov2011}. Due to the explosive nature of the generated plasmonic nanobubbles, intense acoustic waves are emitted, an effect which may be used in the destruction of biological cellular components at the local scale~\cite{lin1998,faraggi2005,wen2009,lombard2021}.

It is generally believed that boiling around hot colloidal nanoparticles is a spinodal process taking place when the fluid is locally brought to its spinodal temperature
~\cite{lowen1992,caupin2006,hou2015,sasikumar2014,sasikumar2014-2,pu2020,maheshwari2018}.
This conclusion has been drawn either from experimental investigations~\cite{siems2011,carlson2012,hou2015} or from molecular dynamics calculations~(MD)~\cite{sasikumar2014,sasikumar2014-2,maheshwari2018,pu2020}. In this scenario, the wetting properties of the solid surface play a minor role and vapor formation is expected to occur homogeneously around the nanoparticle surface. 
Deviations from the spinodal temperature are either interpreted as the crossing of the kinetic spinodal temperature, or due to the presence of dissolved gas which are known to lower the nucleation energy barrier~\cite{alaulamie2016,wang2018,baffou2014}.

In this work, we demonstrate that nanoscale boiling may occur well below the kinetic spinodal temperature. We reach this conclusion by employing a model of a metallic nanoparticle of tunable wettability immersed in water, and a model of solid surface displaying nanoscale contact angle heterogeneities. Compared to previous simulation works, we consider here a realistic model of water. We highlight the heterogeneous nature of nanoscale boiling and its strong dependence on wetting. In particular, the onset of boiling is shown to be $\sim 100$ K lower than water spinodal temperature in the case of weak wetting~(contact angle around $70^o$). Additionally, nanoscale boiling is shown to be a relatively slow process whose kinetics is controlled by the time-dependent interface thermal conductance at the interface of metal and water. This conductance may drop by one order of magnitude depending on wetting, yielding nucleation times much longer than heat diffusion times.

\begin{figure} [!ht]
\centering
\includegraphics [width=2.5cm,height=2.5cm] {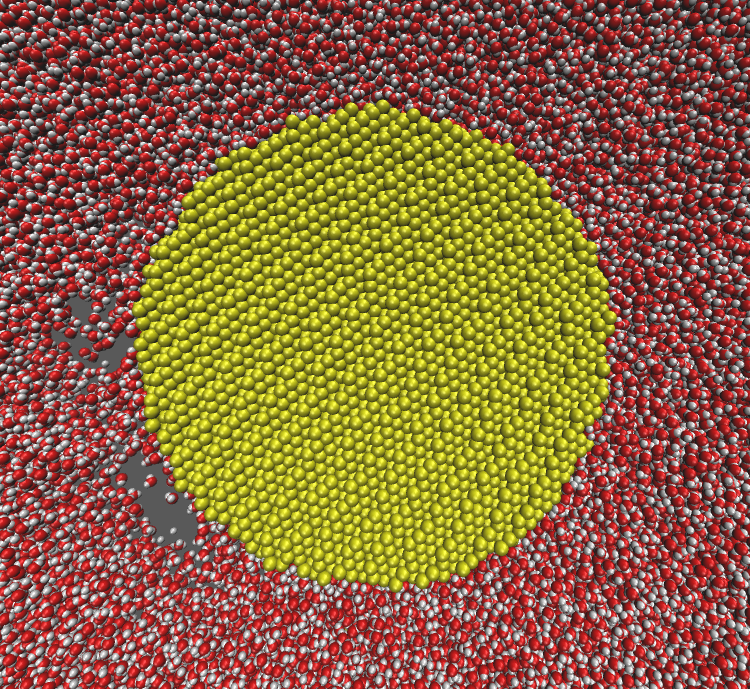}
\includegraphics [width=2.5cm,height=2.5cm] {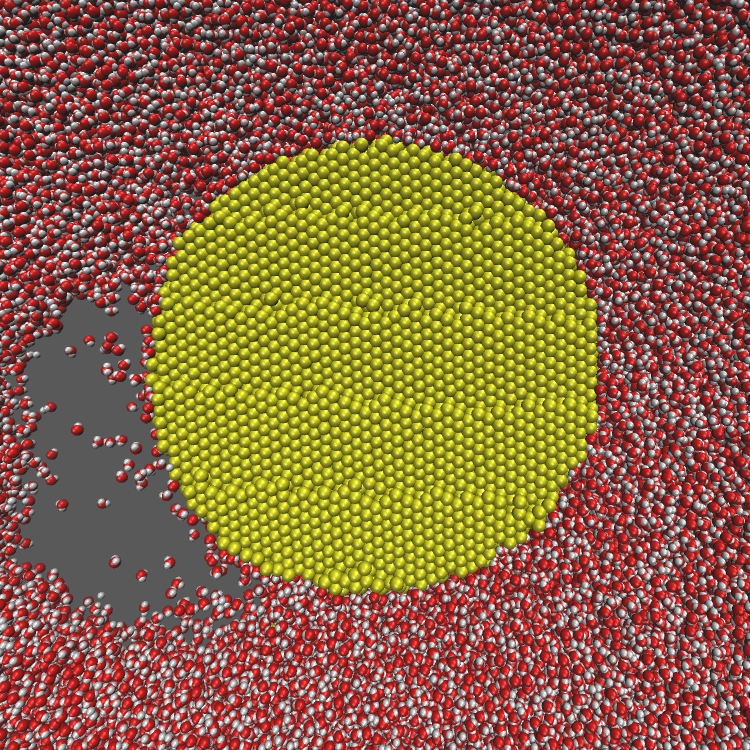}
\includegraphics [width=2.5cm,height=2.5cm] {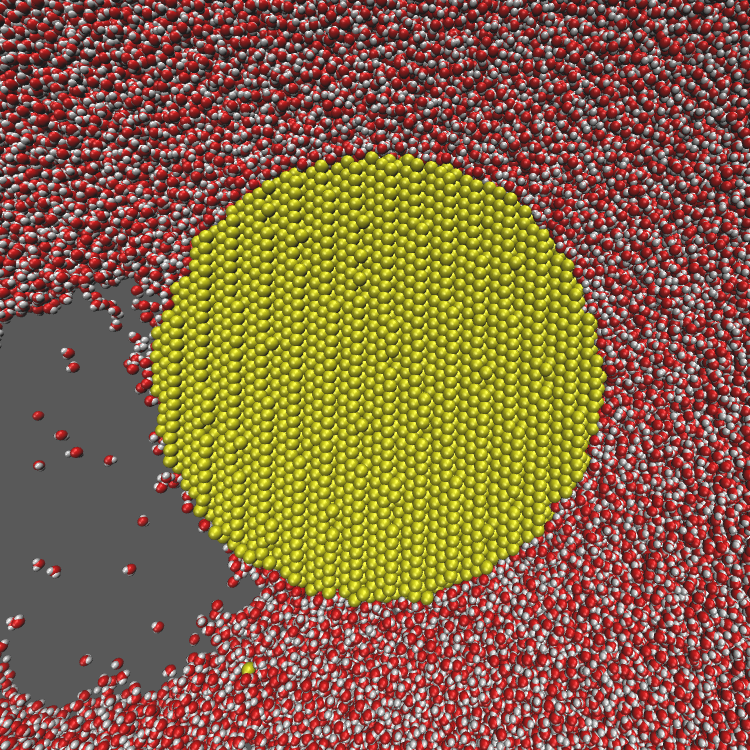}
\caption{Snapshots of the early stages of cavitation for the strong wetting nanoparticle. The sequence of snapshots are taken every $15$ ps at increasing time from left to right. Snapshots for the intermediate and the weak wetting situations display similar behaviors, as shown in the Supplementary Material~\cite{supplementary-material}.}
\label {fig:early-stage-cavitation}
\end {figure}

MD simulations are carried out using the free software Large-scale Atomic/Molecular Massively Parallel Simulator (LAMMPS). More information on the simulation details may be found in the Supplementary material~\cite{supplementary-material}. The nanoparticle consists of a metallic spherical particle immersed in liquid water with fixed diameter of $11$ nm, and occupies a volume fraction $\Phi \sim 10$ \% of the system,  as shown in Fig. \ref{fig:early-stage-cavitation}. Periodic boundary conditions are applied in all directions. The metal-metal interactions are described by the Heinz potential~\cite{heinz2008} while water is described by the flexible TIP4P/2005 \cite{abascal2005} model.  The interaction between the water molecules and the nanoparticle atoms are simulated by a Lennard-Jones (LJ) 12-6 potential, whose strength $\epsilon_{\rm{Au-O}}$ controls the nanoparticle contact angle. Three different wetting regimes are considered: i) strong wetting, ii) intermediate wetting, and iii) weak wetting. The contact angles characterizing the corresponding planar metal/water interfaces are $19^o$, $39^o$, and $71^o$, respectively.  The procedure to measure the contact angles is described in the Supplementary material~\cite{supplementary-material}.
The time step of the dynamics is $1$ fs. The systems are first thermalized in the NPT ensemble using the Nose-Hoover thermostat-barostat for $300$ ps, with a thermostat temperature of $300$ K, and a pressure of one atmosphere~\cite{santamaria2022}. After thermalization, isobaric dynamics are carried out in the NPH ensemble using the Nose-Hoover barostat.
Because in experiments, typically a laser pulse heats the nanoparticles, we consider pulses heating the nanoparticle by velocity rescaling every 40 ps. The target temperature is fixed in each simulation to a value between $2000$ and $3500$ K. 
Simultaneously, the temperature of the water at a distance larger than $95$ \AA~ from the nanoparticle center is rescaled to mantain the heat sink temperature at 300 K. Between each temperature rescaling, the system is allowed to relax in the NPH ensemble.

\begin {figure}
\centering
\subfloat []
{
\includegraphics [width=7.0cm] {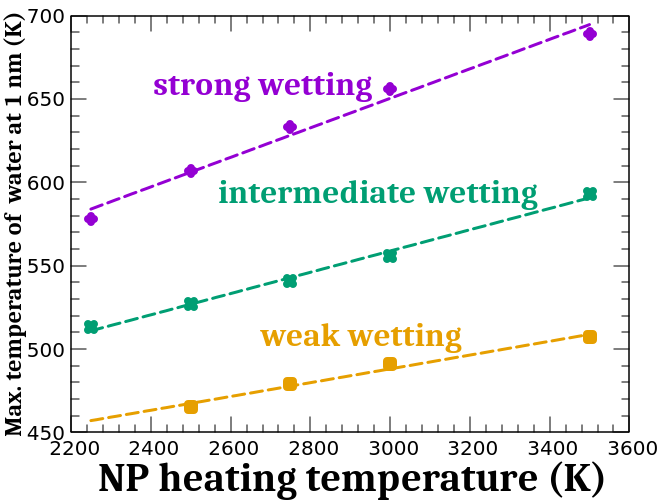}
\label {fig:water-temperatures}
}
\\
\subfloat []
{
\includegraphics [width=7.0cm] {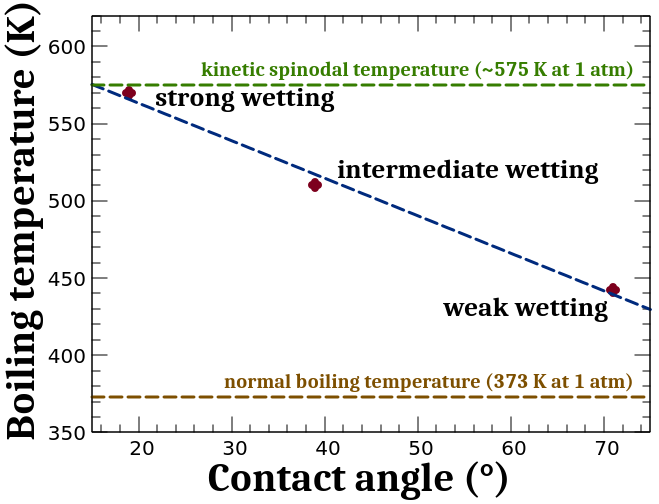}
\label {fig:minimal-temperature-cavitation}
}
\\
\subfloat []
{
\includegraphics [width=7.0cm] {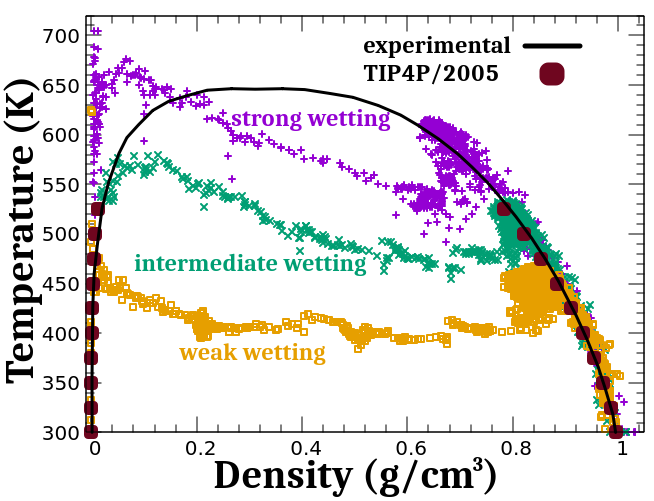}
\label {fig:thermodynamic-state}
}
\caption {a) Maximum temperature reached prior to nucleation by water at a distance $1$ nm from the nanoparticle surface as a function of the nanoparticle heating temperature. b) Minimum water temperature $T_{\rm b}(\theta)$ needed to observe boiling as a function of the contact angle. This temperature corresponds to water at a distance of $1$ nm from the nanoparticle surface. c) Evolution of the thermodynamic state of water at a distance of $1$ nm from the nanoparticle surface for the nanoparticle heating temperature of $2500$ K. The experimental data are taken from \cite{vega2006} while the TIP4P/2005 model saturation curves are extracted from \cite{NIST2018}.}
\label {fig:water-thermodynamics}
\end {figure}
We present now the results obtained when a nanoscale vapor embryo is present in the vicinity of the hot nanoparticle, as illustrated in Fig.~\ref{fig:early-stage-cavitation}. It is first worth noting the heterogeneous nature of early stage boiling. For timescales of a few picoseconds ($<100$ ps), nanobubble nucleation around the heated nanoparticle does not start homogeneously. What is rather observed is that nanobubble generation starts from a random point on the nanoparticle surface. This behaviour is a first indication that nanoscale boiling is not a spinodal process. The heterogeneous nature of boiling is found to be common to all the wetting interactions considered, as seen in the Supplementary Material~\cite{supplementary-material}. 

We confirm the non spinodal nature of nanoscale boiling through a local analysis of water temperature. Fig.~\ref{fig:water-temperatures} shows that the maximal temperature reached by water at a distance $1$ nm from the nanoparticle surface increases linearly with the nanoparticle heating temperature. This behavior is observed for all the wetting regimes, but the temperature reached by the water is higher for the strong wetting nanoparticle. From Fig.~\ref{fig:water-temperatures}, one can identify an onset temperature $T_{\rm b}(\theta)$ which is the minimal temperature to observe boiling. 
Fig. \ref{fig:minimal-temperature-cavitation} shows this onset temperature,  $T_{\rm b}(\theta)$, as a function of the contact angle. The temperature corresponds to the maximum value reached prior to phase change by the water at $1$ nm from the nanoparticle surface. More information on the measurement of this temperature can be found in the Supplementary Material~\cite{supplementary-material}. The onset temperature strongly depends on the nanoparticle contact angle. The weaker the interfacial interaction, the lower the nucleation temperature. This onset temperature is found to be below water spinodal temperature $\sim 573$ K \cite{thiery2010,eberhart2009}. This contrasts with the general belief that nanobubble generation coincides with the crossing of spinodal temperature. Note that the deviations may be large and exceed $100$ K for the weak wetting nanoparticle considered here.

It is also important to emphasize that the cavitation temperatures measured here may be well below the kinetic spinodal temperature of water $T_K$. The kinetic spinodal temperature, which accounts for the possibility to nucleate vapor by thermal fluctuations of the system, is the effective maximal temperature for which a liquid metastable state can be sustained. From experiments~\cite{trinh1978,estefeev1977,estefeev1979,zheng1991,kiselev1999,NIST2018}, we learn that $T_K$ varies typically in the range of $560-578$ K for water. We also estimated this temperature by employing a free energy model which describes accurately the phase coexistence of water~\cite{jeffery1999} and found that $T_K$ is $18$ K below water bulk critical temperature, as detailed in the Supplementary Material~\cite{supplementary-material}. Therefore, the low nucleation temperature that we evidence here are below the water kinetic spinodal temperature. 

Fig.~\ref{fig:thermodynamic-state} shows the evolution of the thermodynamic state of water at a distance of $1$ nm from the nanoparticle surface. We observe that, locally, water follows the saturation line until its temperature reaches the boiling temperature $T_{\rm b}(\theta)$ and phase change occurs. This thermodynamic analysis illustrates also the strong dependence of the boiling temperature on wetting.

\begin {figure} [!ht]
\centering
\subfloat[]
{
\includegraphics [width=7.0cm] {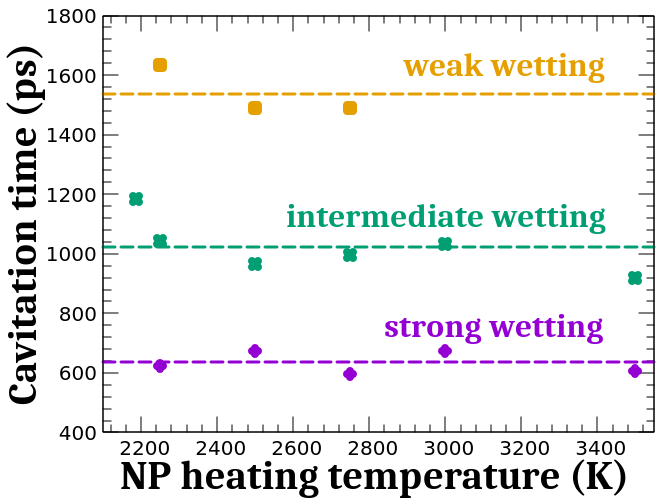}
\label {fig:cavitation-time}
}
\\
\subfloat[]
{
\includegraphics [width=7.0cm] {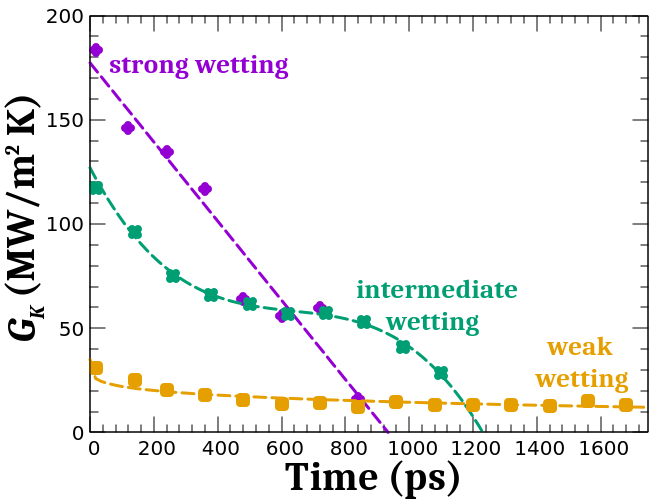}
\label {fig:conductance-variation}
}
\caption{a) Nucleation time as a function of the nanoparticle heating temperature. b) Interfacial thermal conductance $G_{\rm K}$ for the different wetting interactions as a function of the time elapsed prior to boiling. The nanoparticle heating temperature is $T_{\rm np}=2500$ K.}
\label {fig:cavitation-kinetics}
\end {figure}

We turn out to quantify the nanobubble generation kinetics. Fig. \ref{fig:cavitation-kinetics} displays the nucleation time as a function of the nanoparticle heating temperature. This time corresponds to the last instant at which the volume of the system displays negative fluctuations. The time to initiate boiling depends on the nanoparticle contact angle. Strikingly, the cavitation time turns out to be longer for large contact angles, a result which contrasts with isothermal heterogeneous nucleation scenario for which the nucleation energy barrier decreases with the contact angle~\cite{pellegrin2020}. The peculiar kinetics outlined here is related, as we will see, to the ability of the heated nanoparticle to transfer energy to its surrounding fluid. This turns out to be easier for strongly wetting nanoparticles. Surprisingly, the cavitation time is shown to be on average independent of the nanoparticle heating temperature, see Fig.~\ref{fig:cavitation-time}. Nevertheless, the maximum temperature reached by liquid water surrounding the nanoparticle increases with the nanoparticle heating temperature (see Fig. \ref{fig:water-temperatures}).

It is also important to remark that boiling is a relatively slow process. The nucleation time is at least $10$ times larger than the characteristic heat diffusion time over a distance $d=1$ nm, $t_{\rm diff} = {d^2}/{\alpha_{w}} \sim 10 \textrm{ ps} << 1 \textrm{ ns}$, where $\alpha_{\rm w} = 1.5\times 10^{-7}$ m$^2$/s denotes thermal diffusivity of water.  Therefore, our results indicate that the phase change phenomenon observed here is not  diffusion limited. An additional relaxation time $\tau_{\rm np} = C_{\rm p}/ (4\pi R_{\rm np}^2G_{\rm K})$ may characterize the nanoparticle cooling down. In this latter expression, $C_p$ and $R_{\rm np}$ are the heat capacity and the radius of the nanoparticle, respectively and $G_{\rm K}$ the interfacial thermal conductance. The interfacial conductance depends on the contact angle, an effect which has already been outlined~\cite{ge2006,shenogina2009,tascini2017}. More surprisingly, the interface conductance significantly decreases with time prior to water boiling, as reported in Fig.~\ref{fig:conductance-variation}. Further details concerning the calculation of the time-dependent interfacial conductance may be found in the Supplementary material~\cite{supplementary-material}. Note that the conductance drop strongly depends on the nanoparticle contact angle and it is more striking for the strong wetting nanoparticle (around $11$ fold drop) than for the weakly wetting nanoparticle ($2$ fold decrease). This decrease of the interfacial thermal conductance before phase change has been also noticed in experimental studies \cite{kotaidis2006,jollans2019}.

In our case, the conductance drop is responsible for long (ns) cooling times $\tau_{\rm np}$ as evidenced in Fig.~\ref{fig:cavitation-kinetics} and thus plays a leading role in the phase change kinetics. As supported by an analysis presented in the Supplementary material~\cite{supplementary-material}, continuum heat transfer models can describe the MD results only if the variation of the interfacial thermal conductance with time is taken into account. Note that the agreement between MD and continuum heat transfer models is good for strongly and intermediate wetting nanoparticles. Partial disagreement observed for weakly wetting nanoparticles may be due to the presence of a water layer surrounding the nanoparticle with low conductivity and low density. This layer hinders thermal transport at the interface and contributes also to slow down phase change.

\begin {figure} [!ht]
  \centering
\includegraphics [width=6.0cm] {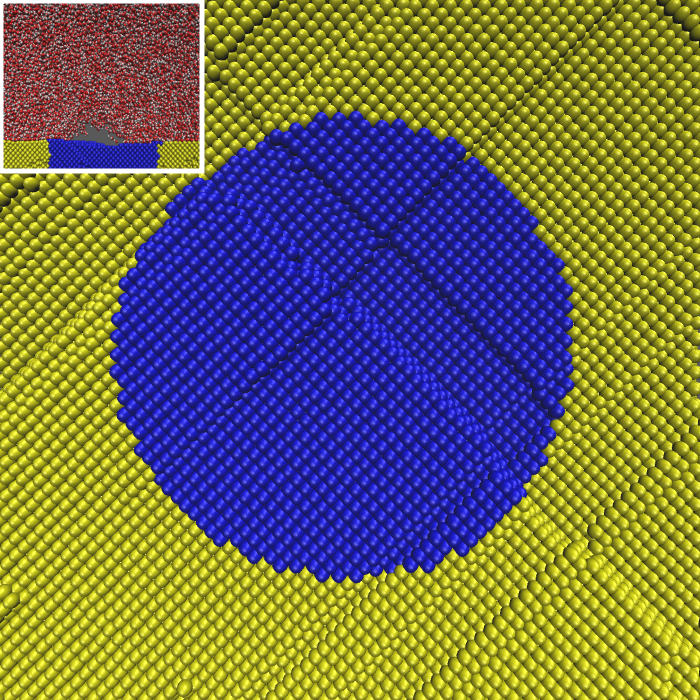}
\includegraphics [width=8.0cm] {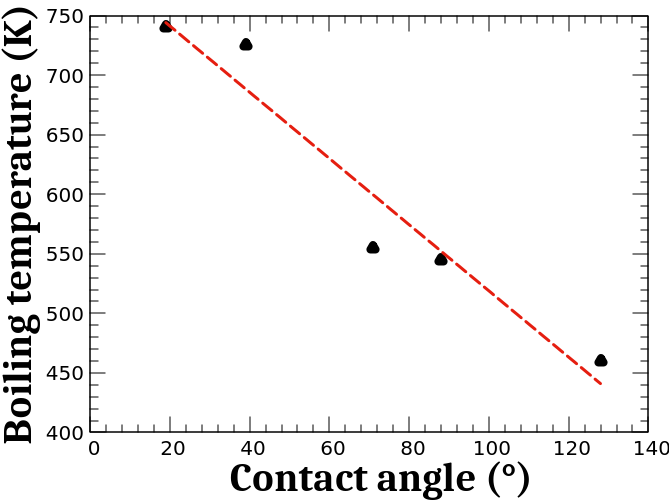}
\caption {Top: top view of a metallic surface having a $11$ nm diameter disk of tunable wettability in its center. The surface is immersed in water. The inset shows a transverse view of the system. The solid slab has a length $15$ nm and $2$ nm height. 
  Bottom~:~Minimum boiling temperature of water above the composite solid surface, as a function of the contact angle of the disk.
  Water boiling temperature is calculated at a distance $1$ nm from the metallic surface. 
}
\label {fig:min-temp-planar}
\end {figure}

So far, we discuss boiling around hot nanoparticles, and it is important to question the general aspect of the conclusions drawn before. To illustrate the generality of the scenario outlined, we consider boiling simulations of a planar solid composite surface in contact with water. The metallic surface is heterogeneous, as can be seen in Fig.~\ref{fig:min-temp-planar}. While most of the solid surface is strongly wetting water~(contact angle $\theta \sim 19^\circ)$, a $11$ nm diameter spherical disk at the center of the surface has a different contact angle. The protocol to heat up the metallic composite surface is the same as the one adopted for the metallic nanoparticle. Figure~\ref{fig:min-temp-planar} displays the boiling temperature of water at a distance $1$ nm from the planar surface as a function of the patch contact angle. Again, we see that the water boiling temperature strongly decreases when the disk is less hydrophilic. This shows that first, contact angle has a strong influence on nanoscale boiling. Secondly, this dependence opens the way to tune the boiling properties of surfaces by designing nanoscale motifs with relatively high contact angles.    

In summary, we show that contrary to the common belief, boiling around heated nanoparticles in water may occur well below the spinodal temperature $\sim 573$ K. Heat transfer phase change is also found to be heterogeneous and starts from the nucleation of a small embryo. Nanoscale boiling kinetics is controlled by the solid/water interfacial thermal conductance, which is found to drop sharply before phase change. This drop explains the relatively long nucleation times observed especially for weak wetting interactions. Beyond nanoparticles, we have shown that the boiling temperature of composites surfaces with nanometer scale heterogeneities changes drastically as a function of the local contact angle. Hence, considering nanoscale patches with different contact angles is a promising strategy to tune the boiling properties of solid surfaces. 

We acknowledge fruitful discussions with F. Caupin, X. Noblin and M. Orrit. All the simulations have been run at the PSMN (P\^ole de Scientifique de Mod\'elisation Num\'erique) of the ENS Lyon. Funding from the IDEX Lyon program, the Van Gogh PHC project nanocalefaction, and the ANR project CASTEX are acknowledged.

\end{document}